\documentstyle[12pt,aasms4]{article}

\begin{document}

\title{Precise Interplanetary Network Localization of the Bursting
Pulsar GRO J1744-28}

\author{K. Hurley}
\affil{University of California, Berkeley, Space Sciences Laboratory,
Berkeley, CA 94720-7450}
\authoremail{khurley@sunspot.ssl.berkeley.edu}

\author{C. Kouveliotou} 
\affil{Universities Space Research Association at NASA Marshall Space Flight Center, 
ES-84, Huntsville AL 35812}

\author{T. Cline}
\affil{NASA Goddard Space Flight Center, Greenbelt, MD 20771}

\author{D. Cole}
\affil{Jet Propulsion Laboratory, MS 169-327, Pasadena, CA 91109}

\author{M. C. Miller}
\affil{University of Maryland, Department of Astronomy, College Park, MD 20742-2421}

\author{A. Harmon, G. Fishman}
\affil{NASA Marshall Space Flight Center, 
ES-84, Huntsville AL 35812}

\author{M. Briggs}
\affil{University of Alabama in Huntsville, AL 35899}

\author{J. van Paradijs\altaffilmark{1}, \altaffilmark{2}University of Alabama in Huntsville, AL 35899}
\altaffiltext{1}{Astronomical Institute `Anton Pannekoek',
University of Amsterdam, The Netherlands}
\altaffiltext{2}{deceased}

\author{J. Kommers, W. Lewin}
\affil{Massachusetts Institute of Technology, Center for Space Research 37-627, 
Cambridge MA 02139}

\begin{abstract}

We analyze 426 observations of the bursting pulsar GRO J1744-28 by \it Ulysses \rm
and BATSE.  Triangulating each burst, and statistically combining the triangulation
annuli, we obtain a 3$\sigma$ error ellipse whose area is 532 sq. arcsec.  The accuracy
of this statistical method has been independently verified with observations of the soft gamma
repeater SGR1900+14.  The ellipse
is fully contained within the 1 $\arcmin$ radius ASCA error circle of the soft X-ray
counterpart, and partially overlaps the 10 $\arcsec$ radius ROSAT error circle of a source which
may also be the soft X-ray counterpart.  A variable source which has been proposed as a
possible
IR counterpart lies at the edge of the 3 $\sigma$ error ellipse, making it unlikely from a purely statistical
point of view to be associated with the bursting pulsar.  

\end{abstract}


\section{Introduction}

The Bursting Pulsar GRO J1744-28 was discovered with the Burst and Transient Source
Experiment (BATSE) aboard the \it Compton Gamma-Ray Observatory \rm (GRO) in 1995 December
(Fishman et al. 1995; Kouveliotou et al. 1996a).
Between the discovery date and 1997 April, BATSE detected over 5800 type II bursts 
(i.e., accretion-powered, Lewin et al. 1996) from this source
(Woods et al. 1999), many of which were also detected by instruments aboard
the \it Rossi X-Ray Timing Explorer \rm (RXTE: Giles et al. 1996) and \it Ulysses \rm, 
among others (e.g., KONUS-WIND, Aptekar et al. 1998).  
The initial source localization was a 6 $\arcdeg$ radius error circle
(Fishman et al. 1995).  Triangulation with BATSE and \it Ulysses \rm resulted in a 24 
$\arcmin$ wide annulus which intersected this error circle, and the use of the BATSE
Earth occultation technique reduced the area of the localization further (Hurley et al. 1995).
Observations using the \it Oriented Scintillation Spectrometer Experiment \rm (Kurfess et al.
1995; Strickman et al. 1996), and BATSE observations of a variable, pulsating (467
ms period) quiescent source associated with the bursting source 
(Finger et al. 1996a,b; Paciesas et al. 1996) resulted in a still smaller error box.  
A subsequent RXTE
observation produced an $\approx$ 5 sq. arcminute error box (Swank 1996; Giles et al. 1996).
Within this error box, Frail et al. (1996a,b) found a variable radio source.  Observations
of the region around the radio source position with the \it Advanced Satellite for Cosmology and Astrophysics \rm
(ASCA) revealed a pulsating, bursting X-ray source with the same
467 ms period (Dotani et al. 1996a,b) whose position
was consistent with that of the radio source, but a later ROSAT observation (Kouveliotou
et al. 1996b; Augusteijn et al. 1997) with higher angular resolution found an X-ray source 
within the 1 $\arcmin$ radius ASCA error circle which was significantly displaced from the radio position.  The radius of the ROSAT error circle is 10 $\arcsec$, corresponding to a 5 $\arcsec$ statistical
error and a 8 $\arcsec$ systematic error, summed in quadrature.  No confidence
level can be quoted for the systematic error, but the statistical error corresponds
to $\sim \rm 10 \sigma$ (J. Greiner, private communication).

Although the radio source was rejected as a possible counterpart to GRO J1744-28, optical
and near-infrared observations of the ROSAT source region did uncover an object at the
limit of the 10 $\arcsec$ radius ROSAT error circle which appeared to be variable (Augusteijn
et al. 1997; Cole et al. 1997).  These observations were carried out at the European
Southern Observatory (ESO) and at the Astrophysical Research Consortium's Apache
Point Observatory (APO).  In some of the observations, it was not
possible to rule out the apparent detection as an instrumental artifact (Augusteijn
et al. 1997); in others, however, there was no reason to suspect that the detection was
not valid (Cole et al. 1997).

It has been proposed that GRO J1744-28 is a low-mass X-ray binary system (LMXB),
in which a neutron star with a dipole field B $\rm \lesssim 10^{11} G$ accretes
matter from its companion.  The rotation period of the neutron star is 467
ms, the orbital period of the system is 11.8 d, and the system is viewed
nearly face-on (e.g. Daumerie et al. 1996).  The distance is approximately that
of the Galactic center.

Because of the difficulty of identifying the counterpart at various wavelengths
in a crowded region of the sky towards the Galactic center, it is important
to consider the details of the ROSAT observation.  It was a short one (820 s) with the High Resolution Imager (HRI); only 273 photons
were collected, and, in contrast to the ASCA observation, neither pulsations nor bursts were 
detected.  (During the observation,
no bursts were recorded by BATSE or \it Ulysses \rm either, 
and the upper limit to the ROSAT pulsed flux
is consistent with that derived from BATSE and RXTE observations.)  From earlier ROSAT observations
in which the source was not detected, it was concluded that the object was transient;
based on the statistics of transient sources in the galactic plane, it was estimated
that the probability of observing a random source unrelated to the bursting pulsar
was less than 10$^{-4}$.  Since no energy spectra are recorded by the HRI, the ASCA
spectrum was assumed to calculate the source flux; it was found to be $\rm \approx
2 \times 10^{-9} erg \: cm^{-2} s^{-1}$ (unabsorbed) in the 0.1 - 2.4 keV energy range (Augusteijn
et al. 1997).  This observation took place in 1996 March.  For comparison, the fluxes
measured by ASCA in the 2 - 10 keV energy range were  $\rm 2 \times 10^{-8} erg \: cm^{-2} s^{-1}$
in 1996 February and $\rm 5 \times 10^{-9} erg \: cm^{-2} s^{-1}$ in 1997 March (Nishiuchi
et al. 1999).  These fluxes would convert to unabsorbed 0.1-2.4 keV fluxes of
$\rm 9.7 \times 10^{-9} erg \: cm^{-2} s^{-1}$
and $\rm 2.4 \times 10^{-9} erg \: cm^{-2} s^{-1}$ respectively using a simple
extrapolation of the power-law continuum measured by Nishiuchi et al. (1999).

To summarize, there are good arguments both in favor of and against the idea that
the true X-ray and optical/IR counterparts to GRO J1744-28 have been identified.  In favor:

1. this was the only ROSAT source detected within the ASCA error circle,

2. it was transient, and

3. the variable optical/IR source was reliably detected in some of the observations of 
Cole et al. (1997).

Against:

1. no bursts or pulsations were observed by ROSAT (although the short duration
of the observation may be to blame),

2. Augusteijn et al. (1997) estimate that the proposed optical/IR counterpart,
if real, exhibited a change in
its IR flux by a factor of 10 over a period of minutes, 
with no accompanying X-radiation; the detection could have been an artifact, and

3. the proposed optical/IR counterpart lies at the edge of or just outside
the ROSAT error circle (depending on the astrometry).

Here we adopt the view that
the true counterpart to GRO J1744-28 may not yet have been identified 
and localized with certainty, 
and we analyze the observations of bursts from GRO J1744-28 by \it Ulysses \rm
and BATSE in order to better constrain the position of the source.

\section{Observations}

We began this analysis by examining \it Ulysses \rm GRB experiment (Hurley et al. 1992) 
data for each BATSE burst.
Knowing the arrival time of a burst at BATSE, the coordinates of the \it Ulysses \rm spacecraft, and the approximate source
position, we extracted data for $\sim \pm$ 100 s about the \it Ulysses \rm
crossing time.  Although the bursting pulsar was a prolific source, it was
not a particularly intense one, and this procedure resulted in the identification
of only $\approx$ 500 bursts in the \it Ulysses \rm data.  Typically, these
were count rate increases in the 3 - 6 $\sigma$ range.  The vast majority of
them were recorded in the untriggered data, which have a time resolution of 0.25 - 2 s,
depending on the telemetry mode.   
We then retained only those bursts which were recorded
by BATSE with 0.064 s time resolution, since these are the ones which can
be cross-correlated with the \it Ulysses \rm time histories with the best
accuracy.  Figure 1 shows one example. The final data set then consisted of 426 bursts,
of which only 5 were recorded by \it Ulysses \rm in triggered (32 ms resolution) data.
The first event in this set was BATSE \# 4042 on 1995 December 19, and the last was
BATSE \# 6085 on 1997 February 2. 

Triangulation of a single burst results in an annulus of possible arrival
directions whose width depends on the vector between the two spacecraft
and the uncertainty in cross-correlating the two time histories (see, e.g.
Hurley et al. 1999a).  As examples, we show the first and last annuli in figure 2.
Their widths are $\rm \approx 0.9 \arcmin \: and \: 3.8 \arcmin \: (1 \sigma)$ respectively, and they intersect at an angle $\rm \approx 37 \arcdeg$, approximately the same angle as the displacement
of the Earth-\it Ulysses \rm vector during the period between the bursts.
In figure 3 we show the distribution
of the 426 annulus half-widths.  The average total width is $\approx 3.2 \arcmin$.
We can predict what the approximate result might be of combining these annuli
statistically.  Two 3.2 $\arcmin$ wide annuli intersecting at an angle of 
37 $\arcdeg$ form a box shaped roughly like a rhombus with diagonals
3.4 $\arcmin$ and 10 $\arcmin$.  (The actual error region will be an ellipse
inscribed in the rhombus, with minor and major axes somewhat smaller than
the diagonals; for the purposes of this simple estimate we ignore this fact and 
base our calculation on the lengths
of the diagonals, which will give us an overestimate of the final error region
size.)  The statistical combination of the 426 annuli 
should therefore be an elliptical error region with minor and major axes 
approximately $\rm 3.4 \arcmin/\sqrt{426}  \, and \, 10 \arcmin/\sqrt{426} $,
or 10 $\arcsec$ and 29 $\arcsec$ respectively.  We show below that these are in
fact close to, but larger than the final dimensions.
 
The statistical method for combining the results of multiple triangulations
has been outlined in Hurley et al. (1999b).
It consists of defining a chisquare-distributed variate which is a function of an assumed source position in right ascension and declination, and of the parameters describing the triangulation annuli.  Let 
$\alpha, \delta$ be the right ascension and declination of
the assumed source position, and let $\alpha_i, \delta_i, \theta_i$
be the right ascension, declination, and radius of the ith annulus.
Then the angular distance d$_i$ between the two is given by
\begin{equation}
d_i= \theta_i - \cos^{-1}(\sin(\delta) \sin(\delta_i) +
\cos(\delta) \cos(\delta_i) \cos(\alpha - \alpha_i) )  
\end{equation}.
If the 1 $\sigma$ uncertainty in the annulus width is $\sigma_i$, then
\begin{equation}
\chi^{2}=\sum_{i}\frac{d_i^2}{\sigma_i^2}.
\end{equation}
The assumed source position is varied to obtain a minimum chisquare; 1, 2, and 3 $\sigma$
equivalent confidence contours in $\alpha$ and $\delta$ are found by increasing
$\chi^{2}_{min}$ by 2.3, 6.2, and 11.8.

The best fitting position for the 426 annuli is 
$\rm \alpha(2000)=17^h 44^m 32^s, \delta(2000)=
-28^o 44\arcmin 31.7\arcsec$, and has a $\chi^2_{min}$ of 415.7 for 424 degrees of freedom
(426 annuli, minus the two fitting parameters $\alpha, \delta$)\rm.  For a
large number of degrees of freedom m, the $\chi^2$ distribution approaches the
normal distribution with standard deviation $\sqrt{2m}$ and mean m.  Thus
the value we obtain for $\chi^2$ lies 0.27 standard deviations from the mean and
is an acceptable fit.  Figure 4 shows the best fitting position, the ROSAT
and ASCA error circles, and the two slightly different positions for the
proposed optical counterpart found by Augusteijn et al. (1997) and
Cole et al. (1997) (these sources are likely to be one and the
same, considering
their quoted astrometric uncertainties), along with
the 1, 2, and 3 $\sigma$ error ellipses obtained in this analysis.    
The Augusteijn et al. (1997) 
and the Cole et al. (1997) positions for the proposed counterpart lie
at $\chi^2_{min}$+12.3 and $\chi^2_{min}$+15.3, or at the 99.8\% and 99.95\% confidence levels, respectively.  The VLA source position is off the map; it lies at $\chi^2_{min}$+1709,
and is definitely excluded as a candidate in this analysis.
The parameters of the 1, 2, and
3 sigma error ellipses are given in table 1.  Figure 5 shows the distribution
of the distances between the individual annuli and the best fit position.

\section{Accuracy of the Method}

One of the design goals of the \it Ulysses \rm mission was an absolute
timing accuracy of several milliseconds.  To confirm that no large errors
exist in the spacecraft timing and ephemeris, end-to-end timing tests
are routinely carried out, in which commands are sent to the GRB experiment
at precisely known times, and the times of their execution onboard the spacecraft 
are recorded and compared with the expected times.  Because of command buffering
on the spacecraft, there are random delays in the execution of these commands,
and the timing is verified to different accuracies during different tests.
The tests just before, during, and just after the series of 426 bursts analyzed
here took
place on 1995 December 5, 1996 October 1, and 1997 February 19, 
and indicated that the timing errors at those times
could not have exceeded 50, 3, and 1 ms respectively.  For comparison,
the 1 $\sigma$ uncertainties in these triangulations are all greater than
125 ms.  This includes both the statistical errors, and a conservative estimate of
possible unknown timing and spacecraft ephemeris errors.  

Two other independent confirmations of the accuracy of the triangulation
method are first, the 
excellent agreement between the VLA and triangulated positions of SGR1900+14,
using the same statistical method as the one we employ here
(Hurley et al. 1999b), and second,
the agreement between the triangulated positions and the positions of gamma-ray bursts
with optical and/or X-ray counterparts (e.g. Hurley et al. 1999c).  
 
Although there is no reason to suspect timing errors, it is difficult
to prove beyond a doubt that they do not exist, so we have investigated the
effects which such errors would have.  We distinguish between two 
hypothetical types.
The first is a constant, systematic offset in the timing of one spacecraft.  For example,
if the difference in the burst arrival times at the two spacecraft were 
systematically overestimated by a constant value of the order of several
hundred milliseconds for each burst, 
the result would be to increase the radii of all
the annuli, leaving the annulus widths and the coordinates of the annulus
centers unchanged.  (The increase in each radius would be almost, but not
exactly the same, since it depends on the value of
the interspacecraft vector, which changes from burst to burst as the spacecraft
move.)  The new annuli would still be consistent with a best-fitting
position with an acceptable $\chi^2_{min}$, but the position would shift by 15 $\arcsec$
for every 100 ms of offset.

The second is a random error whose average value is zero, but whose
value for any given burst may take on positive or negative values up to
several hundred milliseconds.  To simulate
the effects of such errors we have added a random number to the difference in the
spacecraft arrival times for each burst; the number is drawn from a Gaussian distribution
with mean zero and standard deviation 100 ms.  The effect of such an error would
again be to change only the radii of all the annuli, but by different amounts whose
average would be zero.  Since the annulus widths are unaffected, the $\chi^2_{min}$
for the best-fitting position increases, but not to the point where it becomes
unacceptable or even suspect. The best-fitting position shifted by 6 $\arcsec$
in this simulation.

Other types of errors can of course be imagined, but we reiterate that there
is neither any indication that such errors exist, nor any means to disprove their
existence entirely.

\section{Discussion and Conclusions}

Because pulsations and bursts were detected during the ASCA observation,
there is no doubt that ASCA detected the X-ray counterpart to GRO J1744-28, 
but it is not well localized.  
If we accept the ROSAT source as the counterpart, then the 
combination of the 3 $\sigma$ error ellipse derived here and the ROSAT error circle
gives a new, smaller error box whose area is $\approx$ 150 sq. arcsec., or
about one half the ROSAT area.  One reason to accept it is the fact that
the error ellipse indeed overlaps it partially; we estimate the chance
probability of an overlap between the two within the ASCA error circle
to be $\sim$0.14.  If we reject the ROSAT source as the counterpart,
the appropriate error box for GRO J1744-28 becomes the
entire 532 sq. arcsec. 3 $\sigma$ error ellipse.  
However, this implies that the X-ray counterpart must have faded to an undetectable flux 
during the ROSAT observation, or $\rm <5\times10^{-12}
erg \, cm^{-2}s^{-1}$ (unabsorbed).  

In either case, the possible variable IR source
is at or beyond the 3 $\sigma$ confidence levels
of both the ROSAT and the triangulation regions.  From a purely
statistical point of view it is unlikely to be
the counterpart, but it cannot be completely ruled out.  The IPN error ellipse
has been examined in four of the archived
K' images taken at APO and ESO.
Their dates and limiting magnitudes are 1996 January 21 (APO: 14.4 $\pm$ 0.3), 
1996 January 30 (APO: 15.2 $\pm$ 0.3), 1996 February 8 (ESO: 16.75 $\pm$ 0.3), and
1996 May 1996 (ESO: 17.1 $\pm$ 0.3).  Comparing the first three with the last reveals
no variable objects other than the previously identified IR source.  
However, based on the magnitudes of
LMXB's, Augusteijn et al. (1997) estimated that the quiescent counterpart to
GRO J1744-28 might have a K magnitude $\approx 18.7$, or at least two magnitudes
fainter than the completeness limit of their observations, and  Cole et al. (1997)
estimated that observations down to K'=20 were needed.  It is also possible
that the true counterpart is considerably farther away than the Galactic center, or
that absorption in this direction is greater than expected.  

Fortunately, it may be possible to resolve the ambiguity.  The X-ray counterpart
can be detected in an observation
with the \it Chandra \rm High Resolution Camera (HRC) if its flux has not decreased
by more than a few orders of magnitude.  Detection of pulsations would
lead to an unambiguous identification of the counterpart, and the  1 $\arcsec$
HRC resolution would provide the smaller error box needed to carry out deeper searches for
the IR counterpart.

\acknowledgments
KH is grateful to JPL for \it Ulysses \rm support under Contract 958056,
and to NASA for Compton Gamma-Ray Observatory support under
grant NAG 5-3811.

\newpage

\figcaption{\it Ulysses \rm (red) and BATSE (black) time histories for
trigger \#4317.  The \it Ulysses \rm time resolution is 0.5 s, and the
data are for the 25-150 keV energy range.  The BATSE time resolution is
0.064 s, and the data are for the 25-100 keV energy range.  The time
histories are aligned for the best-fitting lag. \label{fig1}}

\figcaption{Triangulation annuli for the first and last bursts in this study.
The first annulus, for BATSE \# 4042 on 1995 December 19, is the narrower one;
its width is $\rm \approx 0.9 \arcmin \: (1 \sigma)$.  The last annulus is for
BATSE \# 6085 on 1997 February 2; its width is $\: 3.8 \arcmin \: (1 \sigma)$.  \label{fig2}
}

\figcaption{The distribution
of the 426 annulus half-widths.  The average is $\approx 1.6 \arcmin$. \label{fig3}
}

\figcaption{An $\sim \rm 1 \arcmin \times 1 \arcmin$ square region
containing the best fitting position for the statistical combination
of the 426 annuli.  The 1, 2, and 3 $\sigma$ error ellipses surround
this position.  The  10 $\arcsec$ radius ROSAT error circle is also
shown.  The center of the 1 $\arcmin$ ASCA error circle is marked; part of the
circle is visible in the lower left hand corner.  The two slightly different positions for
the proposed optical counterpart found by Augusteijn et al. (1997) and
Cole et al. (1997) are marked ``STAR 1'' and ``STAR 2''. \label{fig4}
}

\figcaption{The distribution of the minimum distances between the 426 annuli
and the best-fit position $\alpha_{bf}, \delta_{bf}$.  The minimum distance for
annulus i is given by
$\rm d_i= \theta_i - \cos^{-1}(\sin(\delta_{bf}) \sin(\delta_i) +
\cos(\delta_{bf}) \cos(\delta_i) \cos(\alpha_{bf} - \alpha_i) ) $
, where $\alpha_i, \delta_i, \rm and \, \theta_i$
are the right ascension, declination, and radius of the ith annulus. 
The distances have been normalized to the annulus widths $\sigma_i$.
For comparison, a Gaussian is plotted with mean zero, standard deviation
unity, normalized to the area under the histogram.\label{fig5}
}

\clearpage

\begin{deluxetable}{cccc}
\tablecaption{\it Parameters of the 1, 2, and 3 $\sigma$ error ellipses.}
\tablehead{
\colhead{Ellipse} & \colhead{Minor axis, arcseconds} & \colhead{Major axis, arcseconds} 
& \colhead{Area, square arcseconds} \\
}

\startdata

1 $\sigma$ 	&	6.4   & 20.7 & 104 	\nl
2 $\sigma$	&	10.5	& 34.0 & 279	\nl
3 $\sigma$	&	14.4	& 46.9 & 532 	\nl

\enddata

\end{deluxetable}


\clearpage

\begin{references}

\reference{}Aptekar, R. et al. 1998, \apj \, 493, 404

\reference{}Augusteijn, T. et al. 1997, \apj \, 486, 1013

\reference{}Cole, D. et al. 1997, \apj \, 480, 377

\reference{}Daumerie, P., Kolgera, V., Lamb, F., and Psaltis, D. 1996, \nat \, 382, 141

\reference{}Dotani, T., Ueda, Y., Ishida, M., Nagase, F., Inoue, H., and Saitoh, Y. 1996a, 
\iaucirc \, 6337

\reference{}Dotani, T., Ueda, Y., Nagase, F., Inoue, H., and Kouveliotou, C. 1996b,
\iaucirc \, 6368 

\reference{}Finger, M., Wilson, R., Harmon, B., Hagedon, K., and Prince, T. 1996a, 
\iaucirc \, 6285

\reference{}Finger, M., Koh, D., Nelson, R., Prince, T., Vaughan, B., and Wilson, R. 1996b, 
\nat \, 381, 291

\reference{}Fishman, G., Kouveliotou, C., van Paradijs, J., Harmon, B., Paciesas, W., Briggs, M., Kommers, J., and Lewin, W. 1995, \iaucirc \, 6272 

\reference{}Frail, D., Kouveliotou, C., van Paradijs, J., and Rutledge, R. 1996a, \iaucirc \, 6307

\reference{}Frail, D., Kouveliotou, C., van Paradijs, J., and Rutledge, R. 1996b, \iaucirc \, 6323

\reference{}Giles, A., Swank, J., Jahoda, K., Zhang, W., Strohmayer, T., Stark, M., 
and Morgan, E. 1996, \apj \, 469, L25

\reference{}Hurley, K. et al. 1992, Astron. Astrophys. Suppl. Ser. 92, 401

\reference{}Hurley, K., Kouveliotou, C., Harmon, A., Fishman, G., Briggs, M., van Paradijs, J., Kommers, J., and Lewin, W. 1995, \iaucirc \, 6275

\reference{}Hurley, K., Briggs, M., Kippen, R., Kouveliotou, C., Meegan, C., Fishman, G., Cline, T., and Boer, M. 1999a, \apjs \, 120, 399

\reference{}Hurley, K., Kouveliotou, C., Cline, T., Mazets, E., Golenetskii, S., Frederiks, D., and van Paradijs, J. 1999b, \apj \, 523, L37

\reference{}Hurley, K. et al. 1999c, \apj \, in press (astro-ph 9907180)

\reference{}Kouveliotou, C., van Paradijs, J., Fishman, G., Briggs, M., Kommers, J., Harmon, B., Meegan, C., and Lewin, W. 1996a, \nat \, 379, 799

\reference{}Kouveliotou, C., Greiner, J., van Paradijs, J., Fishman, G., Lewin, W., Rutledge, R., Kommers, J., and Briggs, M. 1996b, \iaucirc \, 6369

\reference{}Kurfess, J., Grove, J., and Tueller, J. 1995, \iaucirc \, 6276

\reference{}Lewin, W., Rutledge, R., Kommers, J., van Paradijs, J., and Kouveliotou, C.,
1996, \apj \, 462, L39

\reference{}Nishiuchi, M. et al. 1999, \apj \, 517, 436

\reference{}Paciesas, W., Harmon, B., Fishman, G., Zhang, S., and Robinson, C. 1996, 
\iaucirc \, 6284

\reference{}Strickman, M. et al. 1996, \apj \, 464, L131

\reference{}Swank, J. 1996, \iaucirc \, 6291

\reference{}Woods, P. et al. 1999, \apj \, 517, 431

\end{references}
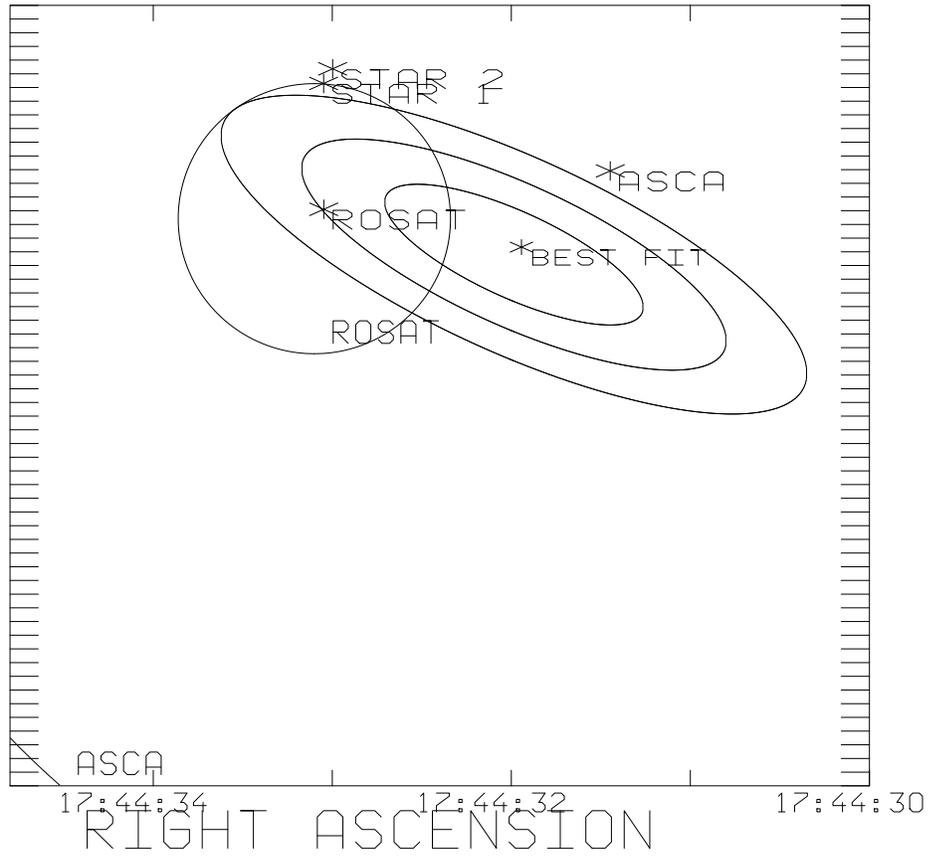
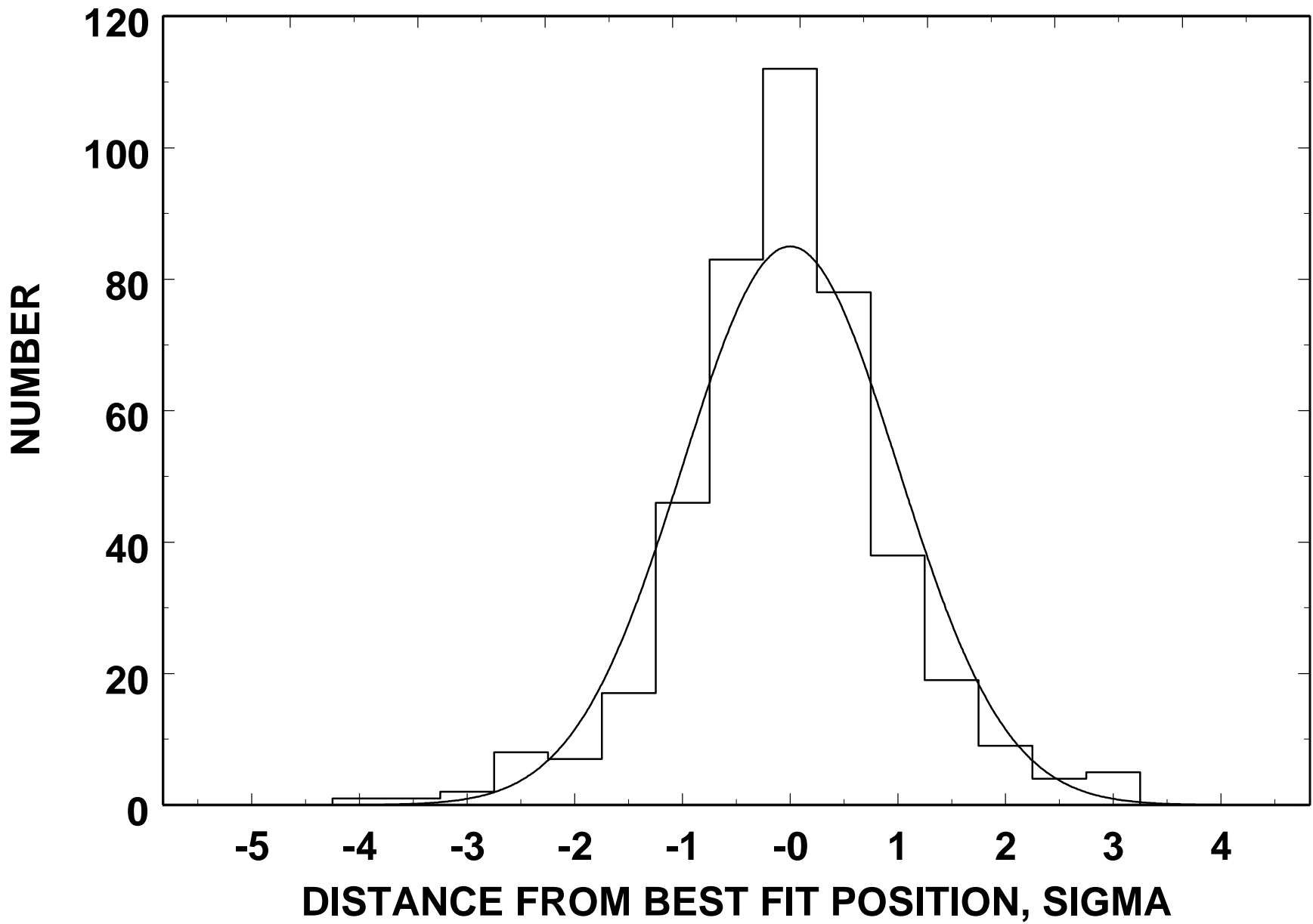
\end{document}